\documentclass[twocolumn,prl,preprintnumbers,amsmath,amssymb,floatfix]{revtex4}
\usepackage[linktocpage,bookmarksopen,bookmarksnumbered]{hyperref}
\usepackage{graphicx}
\usepackage{dcolumn}

\usepackage{amsmath,graphics,epsfig,color,verbatim,ulem}

\begin{document}

\setlength{\pdfpageheight}{\paperheight}
\setlength{\pdfpagewidth}{\paperwidth}

\title{Temperature dependent electronic structures and the negative thermal expansion of $\delta$ Pu}
\author{Z. P. Yin}
\author{Xiaoyu Deng}
\author{K. Basu}
\author{Q. Yin}
\author{G. Kotliar}
\affiliation{Department of Physics and Astronomy, Rutgers University, Piscataway, NJ 08854.}
\date{\today}
\begin{abstract}
We introduce a temperature-dependent parameterization in the modified embedded-atom method and combine it with molecular dynamics to
simulate the diverse physical properties of the $\delta$ and $\varepsilon$ phases of elemental plutonium.
The aim of this temperature-dependent parameterization is to mimic the different magnitudes of correlation strength of the
Pu $5f$ electrons at different temperatures.
Compared to previous temperature independent parameterization,
our approach captures the negative thermal expansion and temperature dependence of the bulk moduli in the $\delta$-phase.
We trace this improvement to a strong softening of phonons near the zone boundary 
and an increase of the $f$-like partial density and anharmonic effects induced by the temperature-dependent parameterization
upon increasing temperature.
Our study suggests it is important to include temperature-dependent parameterization
in classical force field methods to simulate complex materials such as Pu.
\end{abstract}
\maketitle

\section{Introduction}
Plutonium, one of the most complex elemental metals,
has been a subject of intensive theoretical and experimental research
for more than five decades due to its unique anomalous physical properties (see~\cite{LAscience2000} for a review).
Plutonium has six equilibrium solid phases at ambient pressure but different temperatures.
It undergoes a very large volume change (26 percent atomic volume expansion)
during the phase transition going directly from the $\alpha$ phase to the $\delta$ phase
when the latter is stabilized by a small concentration of impurities.
The thermal expansion in the $\delta$ phase of Pu is negative, which is very unusual in metals;
the melting point is very low relative to the other members of the actinide series;
the shear anisotropy of $\delta$-Pu is the largest among all the elements;
and the Gruneisen parameter of $\delta$-plutonium is high compared with those of most elemental metals\cite{Ledbetter}.
Furthermore, the elastic moduli of $\delta$-Pu have a surprisingly strong temperature dependence,
softening with increasing temperature\cite{Migliori}.
The physical origin of the thermal properties, and in particular the negative thermal expansion in the $\delta$ phase,
remains mysterious and has not yet been fully accounted for by first-principles electronic structure methods.
The thermodynamics of elemental plutonium, while having some similarities with those of other heavy fermion materials,
also pose unique challenges. The linear term of the electronic part of the specific heat is substantially enhanced over its local density approximation (LDA) band theory value.
At higher temperatures, the temperature dependence of the electronic specific heat is anomalous.
Different interpretations of this contributions as originating
from the electron-phonon interaction and the electron-electron
interaction have been discussed\cite{Graf:2005, Lashley:2003, Lawson_recent}.
Furthermore the transition from $\alpha$ to $\delta$ phase
is characterized by a large excess of entropy beyond the phonon contribution,
but no comprehensive microscopic theory of the excess entropy is available.\cite{manley1, manley2}

A very useful phenomenology for $\delta$-Pu and its alloys has been constructed by Lawson and his collaborators, by using
analogies with the Invar model of iron \cite{lawson1, lawson2}. In this picture, Pu can be in two states which are associated to very different volumes
and energies, and the free energy can be used to reproduce the elastic and thermodynamic properties of $\delta$ Pu and their dependence on impurities.

As pointed out by Johansson, \cite{Johansson} $\delta$-Pu sits at a special position in the actinide series,
straddling between the so-called \textit{light} actinides (Th to Np, with itinerant $5f$ electrons, smaller atomic volumes and metallic characteristics)
and the \textit{heavy} actinide elements (Am to Es, with larger atomic volumes and insulator-like localized $5f$ orbitals).
The $\delta$-Pu $f$ electrons are thus close to a localization-delocalization transition or crossover,
in which their dual wave-like and particle-like character is important.~\cite{lander} 
This feature also appears clearly in the rearranged periodic table of Smith and Kmetko\cite{smith}.

Modern studies of this material, using a combination of density functional theory and dynamical mean field theory (DFT+DMFT) have substantiated and refined this picture.\cite{Savrasov2001, Dai2003, Toropova:2007, mixed_valence, Marianetti2008}
To describe plutonium, a large number of atomic configurations are required,
and valences with $n_f$=5 and 6 have substantial weight in the ground state.
This multi-determinantal mixed-valent character
has been recently confirmed by resonant $x$-ray emission spectroscopy\cite{booth}.
Unfortunately, anharmonic effects, and other effects which result from the motion of the atoms, are beyond the scope
of current implementations of these first-principles techniques.
They can be investigated using powerful molecular dynamic (MD) methods, once a proper parameterization of the
energy as a function of the atomic positions is posited.
These methods require a parameterization of a classical many body potential.
But with a small number of parameters, in a prescribed functional form,
they can be used to predict successfully a large number of physical properties in a host of $s$ and $p$ band elements~\cite{baskesPRB1992, baskesPRB2003, baskesPRB2006, Jelinek2006}.

In an important series of papers, Baskes and collaborators\cite{baskesPRB2000, baskesPRB2005, baskes2012} used an extension of the highly-successful embedded-atom method (EAM)~\cite{Daw1983, Daw1984},
the modified as MEAM which includes explicitly angular terms and can describe more complex crystal structures~\cite{baskesPRB1992}.
With this method, they described many structural properties of elemental Pu at finite temperature~\cite{baskesPRB2000}.
These calculations however do not reproduced the experimentally observed negative thermal expansion in $\delta$-Pu.

Force field methods such as MEAM can be in principle guided or parameterized using insights from first-principles calculations.
The key part of a force field method is a formula $E[ R_1, R_2, .... , R_n]$ 
which expresses the total free energy in terms of fixed atomic positions of all the atoms
${R_1, R_2, ..... , R_n}$. It can be obtained by tracing out all the electronic coordinates,
in a full path integral treatment of ions and electrons, for a fixed position of the ions. In practice, for weakly correlated systems, it is approximated by
a total energy calculations performed within the Born-Oppenheimer approximation at zero temperature but finite temperature
corrections describing the smearing of the Fermi function can be taken into account\cite{ackland0}.

In this work, we propose a simple modification of Baskes' parameterization of the MEAM force field method. We argue, that
since in $f$-band materials like Pu, where \textit{strong Coulomb repulsions} between electrons results in a strongly temperature
dependent electronic structure, the many-body potential used in the MD simulations has to contain an intrinsic temperature dependence.
This temperature dependence in the form of Kondo resonance, is well established in cerium based materials and is seen in
the DMFT spectral function of many compounds, as for example elemental cerium and the heavy-fermion material $CeIrIn_5$~\cite{shim2007}.
This temperature dependence has also been parameterized recently in the context of the two fluid picture~\cite{pines2007}.
In Pu the characteristic energy scale $T^*$ is estimated to be about $800 K$\cite{Marianetti2008}.
Incorporating this idea into the MEAM parameterization of the free energy, we obtained with a small modification an improved description of
the physical properties of Pu, including the negative thermal expansion of the $\delta$ phase.

\section{Model and Computational Details}

We use the atomistic model of plutonium formulated by Baskes~\cite{baskesPRB2000}, 
except for introducing temperature dependence into the parameters $\beta^{(0)}$ and $\beta^{(3)}$. 
At the heart of the MEAM formalism, the total energy of a system of monatomic atoms is given by
\begin{eqnarray}
E_{\text{tot}} = \sum_i [ F (\rho_{i}) + \frac{1}{2} \sum_{j\neq i} \phi(|r_{ij}|)],
\label{eq:totalenergy}
\end{eqnarray}
where the first term on the right hand side is the embedding function $F(\rho_{i})$,
the energy required to embed an atom into the background host density $\rho_i$ at the lattice site $i$, which is calculated as
\begin{equation}
F(\rho )=AE_{c}\frac{\rho }{\rho _{0}}\ln (\frac{\rho }{\rho _{0}})
\label{}
\end{equation}
where $A$ is an empirical parameter, $\rho_0$ the background density of the reference structure and $E_c$ the cohesive energy.
The second term, the pair potential $\phi$, originating from pairwise electrostatic repulsion 
between atoms at lattice sites $i$ and $j$, is a function of the distance $|r_{ij}|$ between them.
\begin{equation}
\phi (R)=\frac{2}{Z}[E^{u}(R)-F(\rho ^{(0)}(R))]
\label{}
\end{equation}
where $Z$ is the number of first-neighbor atoms in the structure 
and $E^{u}$ is assumed the following form
\begin{eqnarray}
E^{u}(R) &=&-E_{c}(1+a^{\ast }+\delta a^{\ast 3}\frac{r_{e}}{R})e^{-a^{\ast
}} \\
a^{\ast } &=&\alpha (\frac{r_{e}}{R}-1)  \notag
\end{eqnarray}
with $r_e$ the nearest neighbor distance of the reference structure and fitting parameters $a^*$, $\delta$, and $\alpha$.
Background electron density $\rho _{i}$ at site $i$ is modeled as a
combination of angular contributions from neighboring atoms.
\begin{eqnarray}
\rho &=&\rho ^{(0)}\sqrt{1+\Gamma } \\
\Gamma &=&\sum_{l=1}^{3}t^{(l)}(\frac{\rho ^{(l)}}{\rho ^{(0)}})^{2}  \notag
\end{eqnarray}
where $t^{(l)}$ ($l$=1, 2, 3) are fitting parameters,  
and the angular components of electron density $\rho ^{(l)}$ ($l$=0, 1, 2, 3) are calculated as weighted
sum of atomic densities:
\begin{eqnarray}
\rho ^{(0)} &=&\sum_{j(\neq i)}\rho _{a}^{(0)}(R_{ij}) \\
\rho ^{^{2}(1)} &=&\sum_{\alpha }[\sum_{j(\neq i)}x_{ij}^{\alpha }\rho
_{a}^{(1)}(R_{ij})]^{2}  \notag \\
\rho ^{^{2}(2)} &=&\sum_{\alpha ,\beta }[\sum_{j(\neq i)}x_{ij}^{\alpha
}x_{ij}^{\beta }\rho _{a}^{(2)}(R_{ij})]^{2}-\frac{1}{3}[\sum_{j(\neq
i)}\rho _{a}^{(2)}(R_{ij})]  \notag \\
\rho ^{^{2}(3)} &=&\sum_{\alpha ,\beta ,\gamma }[\sum_{j(\neq
i)}x_{ij}^{\alpha }x_{ij}^{\beta }x_{ij}^{\gamma }\rho
_{a}^{(3)}(R_{ij})]^{2}-\frac{3}{5}\sum_{\alpha }[\sum_{j(\neq
i)}x_{ij}^{\alpha }\rho _{a}^{(3)}(R_{ij})]^{2}  \notag
\end{eqnarray}
where $x_{ij}^{\alpha }=R_{ij}^{\alpha }/R_{ij}$, and $\rho _{a}^{(l)}$ are
atomic electron densities, taking an exponential form
\begin{equation}
\rho _{a}^{(l)}(R)=e^{-\beta ^{(l)}(R/r_{e}-1)}
\label{}
\end{equation}
with fitting parameters $\beta^{(l)}$ ($l$=0, 1, 2, 3).
The second term in $\rho ^{(3)}$ is added to make the partial electron
densities orthogonal.

In the case of Pu, the atomistic model is parameterized by twelve semi-empirical MEAM parameters\cite{baskesPRB2000}, 
i.e., $E_c$, $r_e$, $\alpha$, $A$, $\beta^{(l)}$ ($l$=0, 1, 2, 3), $t^{(l)}$ ($l$=1, 2, 3), and $\delta$, 
for which we take the same values as in Ref.\onlinecite{baskesPRB2000} except for two parameters discussed in details below. 
Among these parameters, of particular interests are $\beta^{(l)}$ ($l$=0-3), 
which characterize the exponential decay of $s, p, d, f$-like partial densities 
with the distance from the center of an atom. 
In Ref.\onlinecite{baskesPRB2000}, Baskes connected $\beta^{(l)}$ to the bonding character of electronic orbitals. 
For Pu the most important parameters are $\beta^{(0)}$ which has the strongest effect on the potential energy and $\beta^{(3)}$ 
which can mimic the changes of the $5f$ electrons at different temperatures. 
As the emergence of all the phases of Pu is mainly due to thermal effects 
on electronic correlations of $5f$ electrons and the potential energies of the different phases, 
it is therefore natural to introduce a temperature-dependent parameterization of $\beta^{(3)}(T)$ and $\beta^{(0)}(T)$ 
to capture these effects. 

In this paper, we take $\beta^{(3)}$ as a simple linear function of temperature, $\beta^{(3)}(T) = 9.3422 - \frac{T}{1700}$.
This formulation closely follows the variation of the incoherent quasiparticle peak in $\delta$-Pu,
obtained from DMFT calculation\cite{Marianetti2008}.
For $\beta^{(0)}$, we also choose a simple linear function of temperature below 720 K and fix it for temperature above 720 K,
considering the fact that Pu undergoes the $\delta$-$\varepsilon$ phase transition around 720 K which temperature is also close
to the coherence temperature of Pu\cite{Marianetti2008}.
Specifically we choose $\beta^{(0)}(T) = 1.634 + 0.0014*T$ for T$\le$720 K and $\beta^{(0)}(T)=2.642$ for T$>$720 K.
We show that such a temperature dependent parameterization can account for several experimental observations 
which cannot be described by the previous temperature independent parameterization.

Compared to the previous temperature independent study\cite{baskesPRB2000}, 
we include many more atoms and allow much more time for the system to achieve better thermal-dynamical equilibrium. 
Specifically, we use 6$\times$6$\times$6 supercell (864 atoms) for the $\delta$ phase and 8$\times$8$\times$8 supercell (1024 atoms) 
for the $\varepsilon$ phase. We run the MD simulation for 450 ps to allow the system to arrive at good thermal-dynamical equilibrium and use 
the next 50 ps to compute the average values of relevant physical quantities such as volume. 

\section{Results and Discussion}

Using the proposed temperature-dependent parameterization, we performed the MEAM+MD calculations to compute several physical properties, 
including the volume, bulk moduli, and phonon density of states.

\subsection{Volume}

\begin{figure}[!ht]
\centering{
\includegraphics[width=0.99\linewidth]{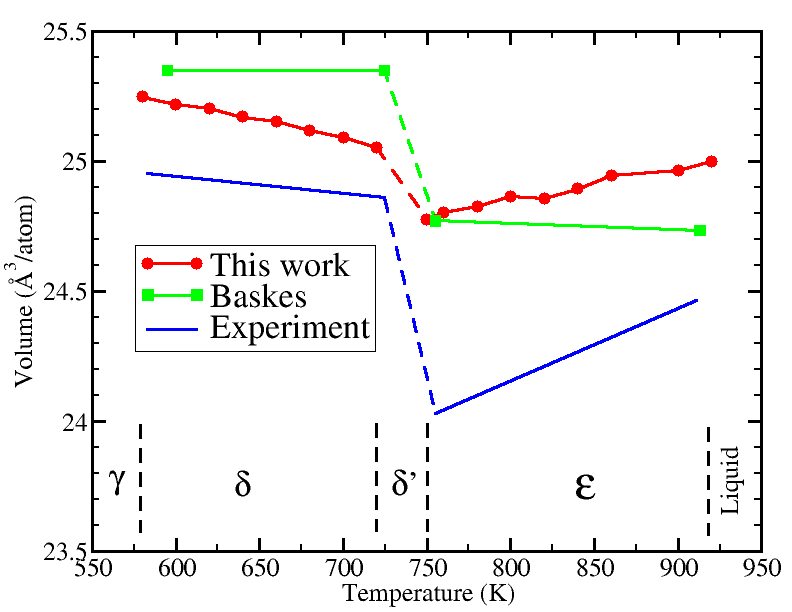}
 }
\caption{
(color online) The calculated volumes of $\delta$ and $\varepsilon$ plutonium at different temperatures using the temperature dependent 
parameterization (this work) and temperature independent parameterization (data taken from Ref.~\onlinecite{baskesPRB2000}) 
and compared with experimental values taken from Ref.~\onlinecite{PuHandbook}.
}
\label{V-T}
\end{figure}

We first show in Fig.\ref{V-T} the calculated volumes of $\delta$ and $\varepsilon$ plutonium at different temperatures 
using the temperature dependent parameterization and compare them with the corresponding results of 
the previous calculations\cite{baskesPRB2000} using temperature independent parameterization. 
Also shown in Fig.\ref{V-T} are the experimental values taken from Ref.~\onlinecite{PuHandbook}.
As mentioned before, an important hallmark of the $\delta$ phase is its negative thermal expansion, 
i.e., its volume decreases with increasing temperature.
This unusual phenomenon doesn't yet receive a satisfactory theoretical explanation.
Clearly, the calculated volume using the temperature independent parameterization has a very weak temperature dependence in the $\delta$ phase, 
and this negative thermal expansion is not captured by the temperature independent parameterization. 
However with our temperature dependent parameterization, we recover the negative thermal expansion in the $\delta$ phase.
This improvement suggests that thermal effects on the $5f$-electronic correlations
must be effectively incorporated into the force field method in order
to recover the phenomenon of negative thermal expansion. 
For the $\varepsilon$ phase, the calculated volume by the temperature independent parameterization is almost flat as a function 
of temperature, 
whereas our temperature dependent parameterization captures the positive thermal expansion in this phase, 
thus improves again previous results of the temperature independent parameterization.

\subsection{Bulk modulus}

\begin{figure}[!ht]
\centering{
\includegraphics[width=0.99\linewidth]{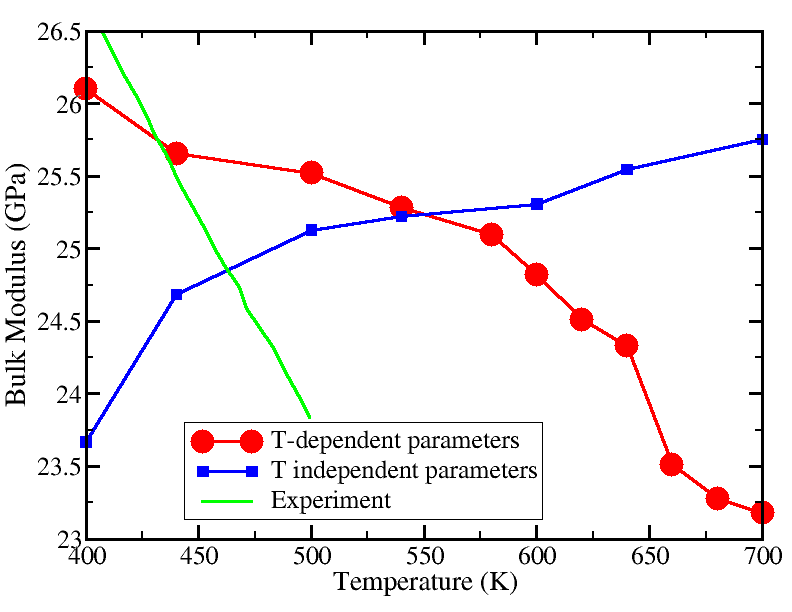}
 }
\caption{
(color online) 
Comparison of the calculated Bulk modulus in $\delta$-Pu using temperature-independent 
and our proposed temperature-dependent parameterizations and with experimental data taken from Ref.\onlinecite{Migliori}.
}
\label{Bulk-Modulous}
\end{figure}

We now proceed to compare the bulk moduli from the temperature dependent and independent parameterizations to experimental studies. 
In recent years, resonant-ultrasound spectroscopy measurements have been able to study the elastic moduli 
of various phases of Pu over a wide range of temperatures \cite{Ledbetter,Migliori,Migliori2007,bourgeois2007}. 
The most remarkable common observation from these experiments is the large elastic softening of the bulk and 
shear moduli with increasing temperature. Such unexpected anharmonic behavior in plutonium has been 
conjectured to arise due to $5f$-electrons undergoing localization-delocalization transition. 
In Figure. \ref{Bulk-Modulous}, we compare the calculated bulk moduli 
using our proposed temperature-dependent parameterization with that using 
traditional temperature-independent one for the $\delta$ phase. 
Experimentally, the bulk modulus was measured up to about $500K$, 
in the stabilized $fcc$ structure \cite{Migliori}, which shows monotonic decreasing $B(T)$ 
as a function of temperature. Although the measurement was not extended to higher temperature, 
the softening tread is expected to continue. In our MD calculations, 
the temperature-dependent parameterization produces the elastic softening with increasing temperature, 
consistent with experiments.  
This comparison shows that the observed softening of 
elastic moduli originates from the $5f$ electronic correlations in Pu, rather than a lattice effect.

\subsection{Phonon density of states}

\begin{figure}[!ht]
\centering{
\includegraphics[width=0.99\linewidth]{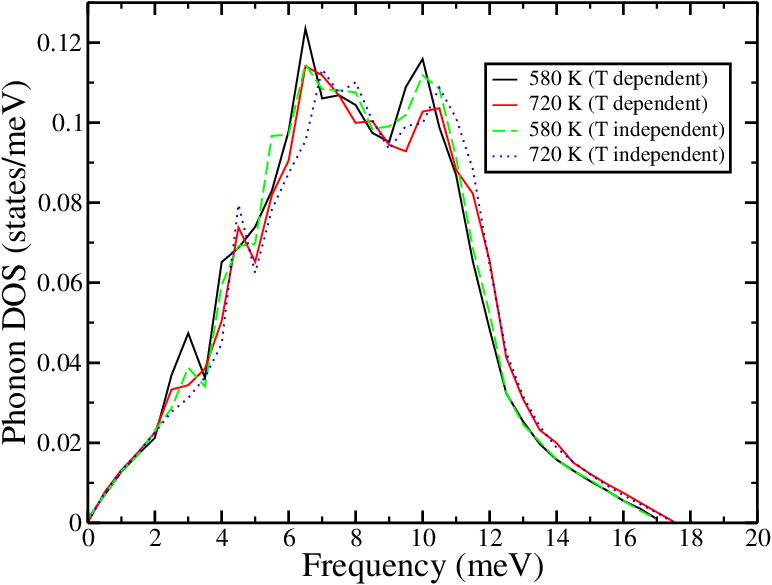}
\includegraphics[width=0.99\linewidth]{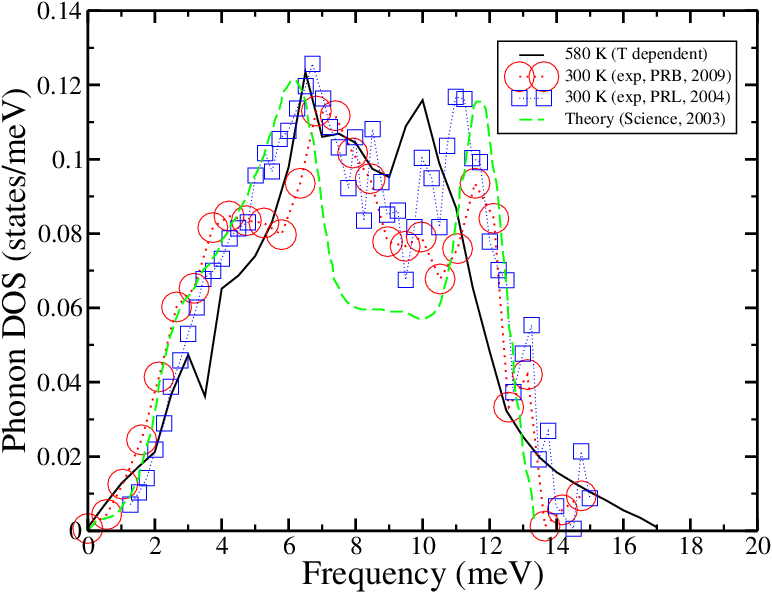}
 }
\caption{
(color online)
Top panel: Calculated phonon density of states in $\delta$-Pu with temperature independent and temperature dependent parameters. 
Bottom panel: Comparison of phonon density of states in $\delta$-Pu with experimental measurements\cite{McQueeney2004, manley1} 
and first-principles theoretical calculations\cite{Dai2003}.
}
\label{phononDOS}
\end{figure}

We also explore the phonon density of states (DOS) in $\delta$ plutonium. 
Experimentally, inelastic neutron scattering (INS) experiments by McQueeney \textit{et al}.~\cite{McQueeney2004, manley1} 
studied the temperature dependence of the phonon spectrum as observed 
from measurements of phonon density of states (DOS). The studies revealed unusual lattice behavior, 
namely a decrease in phonon energies with increasing temperature, 
which again was ascribed to effects of strong electronic correlations of the Pu $5f$ electrons. 
The phonon DOS was not presented in the early MEAM+MD study of Pu~\cite{baskesPRB2000} nor 
was shown in a following lattice vibration study of the $\delta$-Pu~\cite{baskesPRB2005} using the MEAM+MD method. 
Here in the top panel of Fig. \ref{phononDOS}, we present calculated phonon density of states at 
two different temperatures ($580 K$ and $720 K$ ) both with and without 
the temperature-dependent parameterization. 
In the bottom panel of Fig. \ref{phononDOS} we compare our temperature-dependent calculation 
with experimental findings\cite{McQueeney2004, manley1} and previous theoretical work 
of first-principles calculations using DFT+DMFT\cite{Dai2003}. 
Although it is difficult to see the phonon softening effect in our calculated phonon DOS results, 
the main features of our calculated density of states is in good agreement with the measured phonon density of states. 
This agreement verifies the correct physics represented by the MEAM parameters, including our new temperature-dependent ones. 
The MD calculations capture the low frequency peak at around $6.5 meV$, 
which is not visible in the first-principles calculation\cite{Dai2003}. 
However our calculated phonon DOS are slightly off at higher frequencies. 
This discrepancy may be due to temperature effect since our calculations in the $\delta$-Pu are at $580 K$ 
whereas experimental measurements are at room temperature, and in addition the experiments 
have some impurities in the delta-Pu ($5$ percent $Al$ in Ref.~\onlinecite{McQueeney2004} and $2$ percent $Ga$ in Ref.~\onlinecite{manley1})
In the following section, we show that using the MEAM formalism alone without the MD simulations, 
our temperature-dependent parameterization but not the 
temperature independent parameterization captures the phonon softening effects in the $\delta$ phase.

\subsection{Qualitative understanding}

\begin{figure}[!ht]
\centering{
\includegraphics[width=0.99\linewidth]{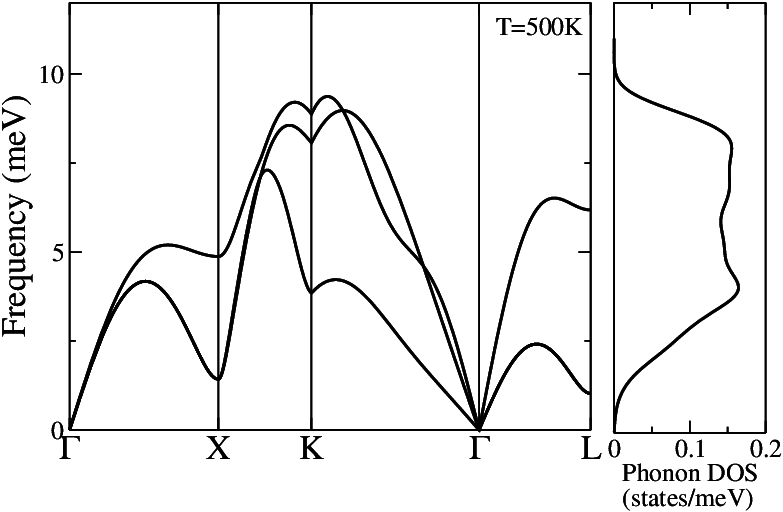}
\includegraphics[width=0.99\linewidth]{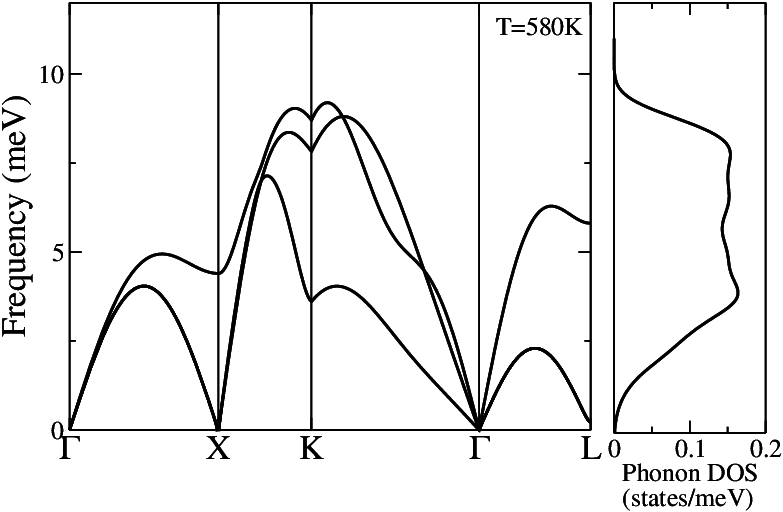}
\includegraphics[width=0.99\linewidth]{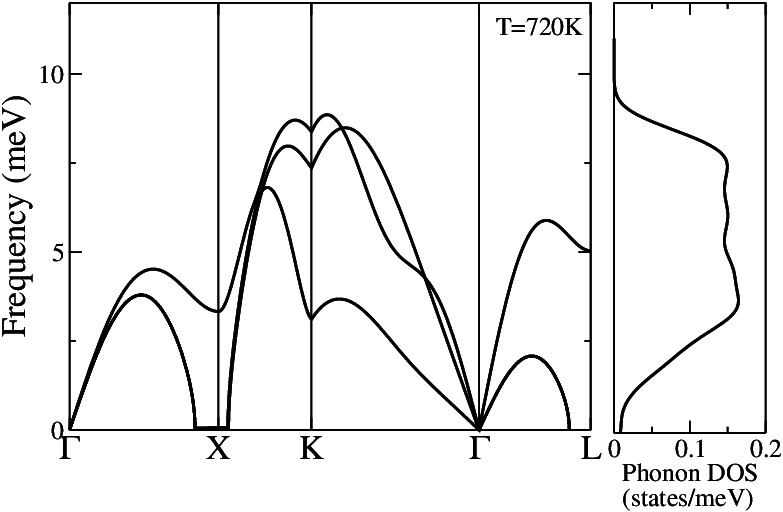}
 }
\caption{
Phonon dispersion and phonon density of states of $\delta$-Pu with lattice constant $a=4.64 \AA$,
calculated analytically from the MEAM model within the harmonic approximation.
The temperature dependence is incorporated by using the temperature dependent $\beta^{(0)}(T)$ and $\beta^{(3)}(T)$.
The phonons near the zone boundary soften strongly upon increasing temperature,
which is consistent with the negative thermal expansion.
}
\label{phonondispersion}
\end{figure}

To understand the physical consequences introduced by the
temperature-dependent MEAM parameters ($\beta^{(0)}$ and $\beta^{(3)}$ in this work),
we have calculated the phonon dispersion of $\delta$-Pu within the harmonic approximation,
where the dynamical matrices are
evaluated solely from the total energy given by the MEAM formalism 
without an explicit temperature dependence as included in the MD simulations. 
Here the temperature dependence
is incorporated by using the temperature dependent $\beta^{(0)}(T)$ and $\beta^{(3)}(T)$.

Our results are shown in Fig~\ref{phonondispersion}. An surprising
observation is that upon increasing temperature, clear and drastic
phonon softenings occur at the $X$ and $L$ points indicating that
the $\delta$-phase tends to be unstable with increasing
temperature, in reminiscent of the unstable phonons of $\varepsilon$-Pu in DMFT calculations\cite{Dai2003} 
by Dai \textit{et al.}.
Meanwhile, the overall scale of the phonon spectra,
phonon density of states and the elastic constants
($C_{11}$,$C_{12}$,$C_{4}$, derived from the long-wave phonon
dispersion) change only slightly, consistent with the results given by
the MD simulations. The long-wave phonon dispersion agrees quite well with
the experimental observation that the $TA$ and $LA$ branches are
nearly degenerated along the $\Gamma$-$X$ direction.\cite{phonon-dispersion}
Also the elastic constants extracted from the dispersion agree well with experimental
results.\cite{elastic-constant} These agreements justify the validity of the current
temperature dependent parameterization. On the other hand, within Baskes's
temperature-independent parameters, the phonon dispersion has no
temperature dependence within the harmonic approximation (the only
possible temperature dependence is from the change of the lattice
constants against the change of temperature, 
however the lattice constant of the temperature independent parameterization has a very weak 
temperature dependence as shown in Fig.\ref{V-T}). Thus, the main effect
introduced by our temperature-dependent parameter is the inclusion of
phonon-softening effects at the level of the harmonic approximation. These
effects seem to be essential to obtain the correct negative thermal
expansion in the $\delta$-phase.

The results also indicate that anharmonic effects induced by
thermal vibrations (molecular dynamics) play an important role
in stabilizing the $\delta$ phase in the MD simulations.
The phonon softening in the harmonic approximation is so drastical
that the $\delta$-phase should be unstable when temperature is higher than 580 K
where the $\delta$-phase is actually stabilized. This implies that the $\delta$-phase is
stabilized by the MD procedure with the help of anharmonic effects in
phonons. This could also be the possible reason why at the zone boundary
the calculated phonon dispersions do not agree well with experimental
results and also why the overall scale of phonon spectra (around 10 meV
in the calculation) is about $20\%$ smaller than in the
experiments\cite{McQueeney2004, manley1} and the MD simulations.

For cubic crystal structures such as the $\delta$ and $\varepsilon$ plutonium,
the anharmonic effects introduced in the MD simulations are reflected in the
non-$s$ angular partial densities $\rho^{(l)}$ ($l=p,d,f$) which are zero in the MEAM methods
and acquire finite values during the MD simulations.
These non-$s$ angular partial densities $\rho^{(l)}$ ($l=p,d,f$) characterizes the 
the weight of the non-spherically symmetric fluctuations in the positions of the neighboring atoms.
The bigger the non-$s$ angular partial densities, the larger the anharmonic effects introduced and
the deviations from the cubic crystal structures. Therefore, the non-$s$ angular partial densities
can also be viewed as a measurement of the randomness of the atoms away from the cubic structures
during the MD simulations.
We find that, compared to the temperature-independent parameterization, the temperature-dependent parameterization 
has more $f$-angular-dependence in the partial charge density in $\delta$ Pu. 
This suggests that it is important to include appropriate amount of $f$-angular-dependent
partial density in $\delta$ Pu in order to account for experimental observations,
and introducing the temperature-dependent $\beta^{(3)}$ and $\beta^{(0)}$ is an efficient way to fulfill this purpose.

While invoking our proposed temperature-dependent parameterization,
we argue that the conventional temperature-independent parameterization misses
some important contributions from the $5f$-electrons in the high temperature (e.g. $\delta$) phases,
and hence it fails to describe certain physical properties (like negative thermal expansion) in the $\delta-$phase,
which arise due to strong correlations of $5f$ electrons. Our temperature-dependent parameterization introduces strong
phonon softening upon increasing temperature and enhances anharmonic effects which help to stabilize the
structure in the MD simulations. A combination of both effects accounts for the unique features of the negative thermal expansion
and the phonon softenings in the $\delta$-Pu.

\section{Conclusion}

In this paper, we have demonstrated that an intrinsic temperature dependent force field method
is needed to simulate the different phases of Pu, especially the $\delta$ phase.
To incorporate this temperature dependence,
we use temperature dependent $\beta^{(3)}$ and $\beta^{(0)}$ parameters in the MEAM+MD calculations.
The negative thermal expansion and the temperature dependence of the bulk modulus in the $\delta$-phase, 
as well as the positive thermal expansion in the $\varepsilon$ phase 
are improved over the original temperature-independent MEAM+MD results and are in reasonable agreements with experiments.
Our temperature dependence parameterization provides the simple idea that the large changes in the spectral function 
of Pu that were shown to take place in the electronic structure calculations\cite{Marianetti2008} should affect the effective 
force between the Pu atoms which are described by the MEAM force field. 
The increase in $\beta^{(0)}$ upon increasing temperature incorporates several important effects compared to the 
temperature independent parameterization,
(1) it introduces strong phonon softenings enhancing anharmonicity; 
(2) it enhances the weight of the non-spherically symmetric fluctuations in the positions of the neighboring atoms;
(3) it lowers the energy of the competing phases which have smaller volumes. Fluctuations into those phases 
decrease the volume in a way reminiscent of the Invar model\cite{lawson1, lawson2},
without having to include explicit Ising degrees of freedom as was done recently by Lee \textit{et al.} 
in Ref.\onlinecite{Lee2012}. 
Our results demonstrate the requirement for finding more accurate atomistic potentials
with temperature-dependent parameterization to incorporate the thermal effects
on the electronic contributions to the system.
To put the parameterization on solid theoretical grounds,
there is the need for a microscopic derivation of the atomistic potentials including the temperature dependence
starting from first-principles electronic structure methods.
Recent advances in the implementations of DFT+DMFT methodologies (see Refs.\onlinecite{Haule-DMFT, Aichhorn-DMFT} for a couple of examples) look very promising in this respect.

\section{acknowledgements}
This work was supported by DOE BES grant number DE-FG02-99ER45761.
K.B. wishes to thank the Alcatel-Lucent foundation for support. 
Q.Y. was supported by the DOE nuclear Energy University Program, contract No. 00088708.
The authors generously thank Mike Baskes and Steve Valone for sharing their MD code and for helpful discussions.

\end{document}